\documentclass{raa}

\usepackage{graphicx,times}             
\input{epsf.sty}                        
\input{psfig.sty}                       

\begin{document}

   \title{A Synchrotron Self-Compton Scenario for the Very High Energy $\gamma$-ray Emission of the Intermediate BL Lacertae
   object W Comae
}

   \volnopage{Vol.0 (200x) No.0, 000--000}      
   \setcounter{page}{1}          

   \author{Jin Zhang\inst{1,2}
   }

   \institute{National Astronomical Observatories/Yunnan Observatory, Chinese Academy of Sciences, Kunming
   650011, China; {\it jinzhang@ynao.ac.cn}\\
        \and
             The Graduate School of Chinese Academy of Sciences\\
        }

   \date{Received~~2009 month day; accepted~~2009~~month day}

\abstract{W Comae has significant variability in
multi-wavelengthes, from the radio to the gamma-ray bands. A
bright outburst in the optical and X-ray bands was observed in
1998, and most recently, a strong TeV flare was detected by
VERITAS in 2008. It is the first TeV intermediate-frequency-peaked
BL Lacertae (IBL) source. I find that both the broadband spectral
energy distributions (SEDs) quasi-simultaneously obtained during
the TeV flare and during the optical/X-ray outburst are well fit
by using a single-zone synchrotron + synchrotron-self-Compton
(SSC) model. The satisfactory fitting requires a large beaming
factor, i.e., $\delta\sim 25$ and $\delta\sim 20$ for the TeV
flare and the optical/X-ray outburst, respectively, suggesting
that both the optical/X-ray outburst and the TeV flare are from a
relativistic jet. The size of the emission region of the TeV flare
is three times larger than that of the optical/X-ray outburst, and
the strength of the magnetic field for the TeV flare is $\sim 14$
times smaller than that of the X-ray/optical outburst, likely
indicating that the region of the TeV flare is more distant from
the core than that of the X-ray/optical outburst. The IC component
of the TeV flare peaks at around 1.3 GeV, but it is around 20 MeV
for the X-ray/optical outburst, lower than that for the TeV flare
with 2 orders of magnitude.  The model predicts that the
optical/X-ray outburst might be accompanied by a strong MeV/GeV
emission, but the TeV flare may be not associated with the
X-ray/optical outburst. The GeV emission is critical to
characterize the SEDs of the optical/X-ray outburst and the TeV
flare. The predicted GeV flux is above the sensitivity of
\emph{Fermi}/LAT, and it could be verified with the observations
by \emph{Fermi}/LAT in near future.
 }

   \authorrunning{Jin. Zhang}            
   \titlerunning{VHE $\gamma$-ray Emission of the IBL W Comae }  

   \maketitle

%
%
\section{Introduction}           
\label{sect:intro} The continuum emission of active galactic
nuclei (AGNs) is both highly luminous and rapidly variable,
especially for a sub-class of blazars. The radiation of blazars is
dominated by emission from a relativistic jet oriented close to
the line of sight (Blandford \& Rees 1978). Their broad band
spectral energy distributions (SEDs) are characterized by two
broad, well separated components. It is well believed that the
lower one is produced by the synchrotron process, and the higher
one could be due to inverse Compton (IC) scattering of the same
electron population (e.g., Ulrich et al. 1997; Urry 1999).
According to the location of the synchrotron hump, blazars are
classified as flat-spectrum radio quasars (FSRQs),
low-frequency-peaked BL Lac objects (LBLs) with a synchrotron
radiation peak at the IR/optical regime, and high-frequency-peaked
BL Lac objects (HBLs) with a synchrotron radiation peak at the
X-ray band (Giommi \& Padovani 1994; Ulrich et al. 1997). The
intermediate-frequency-peaked BL Lac objects (IBLs) fill in the
gap between LBLs and HBLs.

Blazars show significant variability in multi-frequency, from the
radio to the gamma-ray bands, even in the TeV gamma-ray band.
Gamma-ray emission is an important emission component for the SEDs
of blazars, since the radiation in this band is comparable to the
total radiation power of the sources and even higher than that in
the rest of the other energy bands. It plays important roles on
understanding the radiation mechanisms and the emission regions of
Blazars (e.g., Catanese \& Weekes 1999). Considerable samples of
GeV-TeV blazars have been obtained with both the spaced-based
instruments and ground-based Cherenkov telescopes. For examples,
the EGRET instrument on board the Compton Gamma Ray Observatory
(CGRO) identified a sample of some MeV/GeV blazars. Most of them
are FSRQs and LBLs (Hartman et al. 1999; Mattox et al. 2001). The
imaging atmospheric Cherenkov telescopes (IACTs) have established
a sample of more than 20 blazars as very high energy (VHE)
$\gamma$-ray radiation sources. Almost all of them belong to HBLs.
W Comae (W Com, ON 231, 1219+285), an IBL at redshift $z=0.102$,
has been long considered as a VHE gamma-ray source (Kerrick et al.
1995). More recently, a TeV flare of this object was detected with
VERITAS, an array of four imaging atmospheric-Cherenkov telescopes
(Acciari et al. 2008). This is the first TeV detection from an IBL
source. In this paper, I investigate the properties of the VHE
emission from this source.

The historic optical light-curve of W Com shows three major
outbursts peaking in March 1995, February 1996, and January 1997.
The source brightness reached the highest magnitude in April 1998
($R=12.2$ ; Tosti et al. 1999; Massaro et al. 1999), which was
ever observed since 1940\footnote{Wolf (1916) reported $B=11.5$ in
1901 and 1903.}. It was about 3 times brighter than the optical
outbursts observed in the previous years and a factor of 60 than
the minimum brightness $B=17.5$ in 1972 (Tosti et al. 1999;
Massaro et al. 1999). A never detected  high polarization was also
observed at the same time (Tosti et al. 1999), which indicates a
non-thermal origin of the optical outburst. The source was also
detected in infrared band by IRAS, the Infrared Astronomical
Satellite (Impey \& Neugebauer 1988).

In the X-ray band, W Com was observed by \emph{Einstein}/IPC in
June 1980 (Worrall \& Wilkes 1990) and by ROSAT/PSPC in June 1991
(Lamer et al. 1996; Comastri et al. 1997), with detections of a
flux density $\sim 1\mu$Jy and $0.4\mu$Jy at 1 keV, respectively.
The derived energy spectral index from the ROSAT/PSPC observation
is $\alpha\sim 1.2$. The observation with XTE in a
multi-wavelength campaign in February 1996 yielded only an upper
limit (Maisack et al. 1997). The source was in a high state in the
X-ray band during the strongest optical flare in 1998 (Tagliaferri
et al. 2000). The observation with the \emph{BeppoSAX} satellite
derived a well-defined two-component feature in the X-ray
spectrum, which is explained as a synchrotron component and an IC
component (Tagliaferri et al. 2000; B\"{o}ttcher et al. 2002).
Moreover, the source showed a significant variation in the soft
X-ray band (0.1-4 keV), but no similar behavior in the hard X-ray
band (4-10 keV).

In the gamma-ray band, W Com was first observed with CGRO/EGRET
during 1991-1992. The observed gamma-ray spectrum is extremely
hard, with $\alpha\sim0.4\pm0.4$ (Montigny et al. 1995; Sreekumar
et al. 1996), even harder than that derived in 1995 when the
source was in the brightest state in the EGRET band (Tagliaferri
et al. 2000). The EGRET observation in February 1996 during a
quasi-simultaneous multi-wavelength campaign shows that its
brightness was weaker than that observed in November 1993 by a
factor of 1.5 (Maisack et al. 1997). During April 1998 to May 1998
the source was in a high state in both the optical and X-ray
bands, but the gamma-ray emission was only marginally detected by
EGRET in March 1998, with a $2.7 \sigma$ significance level
(B\"{o}ttcher et al. 2002). Observations with Whipple/IACT
obtained some upper limits in the TeV band in 1993/94 (Kerrick et
al. 1995) and 1995/96/99 (Horan et al. 2004). A preliminary
$2\sigma$ upper limit was derived with STACEE in 1998
(B\"{o}ttcher et al. 2002). Interestingly, a strong TeV flare was
detected by VERITAS in the middle of March 2008 (Acciari et al.
2008), but no accompanied ourbursts/flares in the X-ray and
optical bands. An X-ray flare was detected about two weeks after
the TeV flare (Acciari et al. 2008).

As mentioned above, the outburst/flare events of W com in the
optical/X-ray bands and in the TeV band seem not to happen
simultaneously. It is generally believed that the X-ray/optical
emission is produced by the synchrotron radiation. It is a puzzle
if the TeV emission is from the same site as the X-ray/optical
emission through an IC scattering by the same electron population
of the synchrotron radiation. In this paper, I focus on this issue
by fitting the broadband SEDs observed quasi-simultaneously during
the TeV flare in 2008 and during the optical/X-ray outburst phase
in 1998 with a single-zone synchrotron+synchrotron- self-Compton
(SSC) model (Maraschi et al. 1992; Bloom \& Marscher 1996), and
reveal the different properties between the TeV flare and the
X-ray/optical outburst by comparing the fitting results.

\section{Model}
\label{sect:model} The broad band SED for blazars is a
double-peaked structure. It is well believed that the lower one is
produced by the synchrotron process, and the higher one could be
due to inverse Compton (IC) scattering of the same electron
population (e.g., Ulrich et al. 1997; Urry 1999). The IC
scattering photon field could be from the synchrotron radiation
themselves, the so-called SSC model (Maraschi et al. 1992; Bloom
\& Marscher 1996) or from external radiation fields (EC), such as
broad line region (BLR; Sikora et al. 1994; Koratkar et al. 1998),
accretion disk (Dermer et al. 1992), torus (B\L a\.{z}ejowski et
al. 2000), and cosmic microwave background (CMB; Burbidge et al.
1974; Tavecchio et al. 2000). Since those external photon fields
are very weak comparing with the synchrotron radiation field for
the BL Lac objects, I consider only a synchrotron + SSC model by
assuming that the emission region is a homogeneous sphere with
dimension $R$ and the electron distribution in energy is a broken
power law with indices $p_{1}$ and $p_{2}$ below and above the
break energy $\gamma_{b}m_{e}c^{2}$,
\begin{equation}
N(\gamma )=\left\{ \begin{array}{ll}
                    N_{0}\gamma ^{-p_1}  &  \mbox{ $\gamma < \gamma _b$}, \\
            N_{0}\gamma _b^{p_2-p_1} \gamma ^{-p_2}  &  \mbox{ $\gamma > \gamma _b$,}
           \end{array}
       \right.
\end{equation}
where $p_{1,2}=2\alpha_{1,2}+1$, $\alpha_{1,2}$ are the spectral indices, and $\gamma$ is the Lorentz factor
of the electrons. The frequency of the synchrotron radiation is
\begin{equation}
\nu_{syn}=\frac{4}{3}\nu_B\gamma^{2}\frac{\delta}{1+z},
\end{equation}
where $\nu_{B}=2.8\times10^6 B$ Hz is the Larmor frequency in
magnetic field \emph{B} (e.g., Ghisellini et al. 1996) and
$\delta$ is the Doppler factor. The synchrotron emissivity
$\epsilon_s(\nu)$ is calculated with
\begin{equation}
\epsilon_s(\nu)=\frac{1}{4\pi}\int_{\gamma_{min}}^{\gamma_{max}}d
\gamma N(\gamma)P_s(\gamma, \nu).
\end{equation}
Here $P_s(\nu, \gamma)$ is the single electron synchrotron emissivity averaged over an isotropic distribution
of pitch angles. It is calculated with (Crusius \& Schlickeiser 1986; Ghisellini et al. 1988)
\begin{equation}
P_s(\gamma, \nu)=\frac{3\sqrt{3}}{\pi}\frac{\sigma_T cU_B}{\nu_B}
g^2\{ K_{4/3}(g)K_{1/3}(g)-\frac{3}{5}
g[K^2_{4/3}(g)-K^2_{1/3}(g)] \},
\end{equation}
where $g=\nu/(3\gamma^2\nu_B)$, $K_\alpha$ is the modified
$\alpha$-order Bessel function, $\sigma_T$ is the Thomson
cross-section, and $U_B=\frac{B^2}{8\pi}$ is the magnetic field
energy density. The synchrotron radiation field $I_s(\nu)$  is
calculated by the transfer equation,
\begin{equation}
I_s(\nu)=\frac{\epsilon_s(\nu)}{k(\nu)}[1-e^{-k(\nu)R}],
\end{equation}
where $k(\nu)$ is the absorption coefficient (Ghisellini \&
Svensson 1991).

In the SSC scenario, the relativistic electrons interact with
synchrotron radiation photons through the IC scattering. The IC
emissivity is calculated by
\begin{equation}\label{ec}
\epsilon_c(\nu_c)=\frac{\sigma_T}{4}\int_{\nu_i^{min}}^{\nu_i^{max}}
\frac{d\nu_i}{\nu_i}\int_{\gamma_1}^{\gamma_2}\frac{d\gamma}{\gamma^2\beta^2}N(\gamma)f(\nu_i,\nu_c)\frac{\nu_c}{\nu_i}I_s(\nu_i),
\end{equation}
where $\nu_i$ is the frequency of the incident photons emitted by
the synchrotron radiation between $\nu_i^{min}$ and $\nu_i^{max}$,
$\beta=v/c$, $\gamma_1$ and $\gamma_2$ are the lower and upper
limits of the scattering electrons, and $f(\nu_i,\nu_c)$ is the
spectrum produced by scattering monochromatic photons of frequency
$\nu_i$ with a single electron (e.g., Rybicki \& Lightman 1979).
The medium is transparent for the IC radiation field, so I simply
derived $I_c(\nu_c)=\epsilon_c(\nu_c)R$.

Assuming that $I_{s,c}$ is an isotropic radiation field, the
monochromatic luminosity around the source is obtained by
\begin{equation}
L(\delta\nu)=4\pi^2 R^2 I_{s,c}(\nu)\delta^3.
\end{equation}
Then the observed flux density is given by
\begin{equation}
F(\nu_{obs})=\frac{4\pi^2 R^2 I_{s,c}(\nu)\delta^3(1+z)}{4\pi
D^{2}},
\end{equation}
where \emph{D} is the luminosity distance of the source and
$\nu_{obs}=\nu\delta/(1+z)$.

In the GeV-TeV regime, the Klein- Nishina effect could be
significant. It makes the IC spectrum have a high-energy cut-off.
I take this effect into account by using a step function for the
energy dependence of the cross section, $\sigma=\sigma_{T}$ for
$\gamma x\leq 3/4$ and $\sigma=0$ otherwise, where
$x=h\nu/m_{e}c^{2}$ (e.g., Tavecchio, et al. 1998; Chiaberge \&
Ghisellini 1999). Since W com locates at $z=0.102$, the absorption
by the infrared background light and CMB during the gamma-ray
photons propagating to Earth is also considered (Stecker et al.
2006).

\section{Numerical Results}
I fit the broadband SEDs observed quasi-simultaneously during the
TeV flare in 2008 and during an optical/X-ray outburst phase in
1998. The SED data of the optical/X-ray flare are taken from
Tagliaferri et al. (2000, for the optical and the X-ray data) and
B\"{o}ttcher et al. (2002, for the radio and the the EGRET data).
The observed data of the TeV flare in 2008 are from Acciari et al.
(2008) and the references therein. The size of emission region is
estimated by the variability timescale $t$ with $R=ct\delta$,
where $c$ is speed of light. The $t$ value is taken as 10 hours
for the X-ray/optical outburst and 1 day for the TeV flare
according to the timescales of the X-ray flare in 1998 and the TeV
flare in 2008 (B\"{o}ttcher et al. 2002; Acciari et al. 2008).

The well-sampled SED data in the radio, optical, X-ray, and TeV
bands for the 2008 TeV flare place strong constraints on the
physical parameters of the model. Although no TeV detection was
made for the 1998 optical/X-ray outburst, the X-ray spectrum
measured by \emph{BeppoSAX} during the optical/X-ray outburst
reveals a clear two-component feature, i.e., a synchrotron
component and an IC component (Tagliaferri et al. 2000;
B\"{o}ttcher et al. 2002), which also reliably confines the model
parameters. The fitting results are shown in Figure 1, and the
model parameters are reported in Table 1. The SED derived by the
model is corrected for $\gamma\gamma$ absorption according to the
baseline infrared background light case of Stecker et al. (2006).
It is found that the model well fits the SEDs. Large beaming
factors, $\delta\sim 25$ and $\delta\sim 20$ are required to fit
the SEDs, suggesting that both the optical/X-ray outburst and the
TeV flare are from a relativistic jet. It is interesting that size
of the emission region of the TeV flare is three times larger than
that of the optical/X-ray outburst, and the strength of the
magnetic field for the TeV flare is $\sim 14$ times smaller than
that of the X-ray/optical outburst, likely indicating that the
region of the TeV flare is further away from the core than that of
the X-ray/optical outburst. The IC component of the TeV flare
peaks at around $10^{24}$ Hz, but it is around $10^{22}$ Hz for
the X-ray/optical outburst, lower than that for the TeV flare with
2 orders of magnitude.
\begin{table}
\begin{center}
\caption[]{Fitting Parameters}\label{Tab:publ-works}
 \begin{tabular}{c c c c c c  c ccc}
  \hline\noalign{\smallskip}
Epoch & Syn. peak(Hz) &$\alpha_{1}$ & $\alpha_{2}$ & $\delta$ & B (G) & R (cm)\\
  \hline\noalign{\smallskip}
  2008 & $5.2\times10^{14}$ &0.54 & 1.63 & 24.5 & 0.01 & $6.35\times10^{16}$\\
  1998 & $1.0\times10^{14}$ &0.2 & 1.53 & 19.4 & 0.14 & $2.1\times10^{16}$\\
  \noalign{\smallskip}\hline
\end{tabular}
\end{center}
\end{table}
\begin{figure}
   \vspace{0.5cm}
   \begin{center}
   \plotone{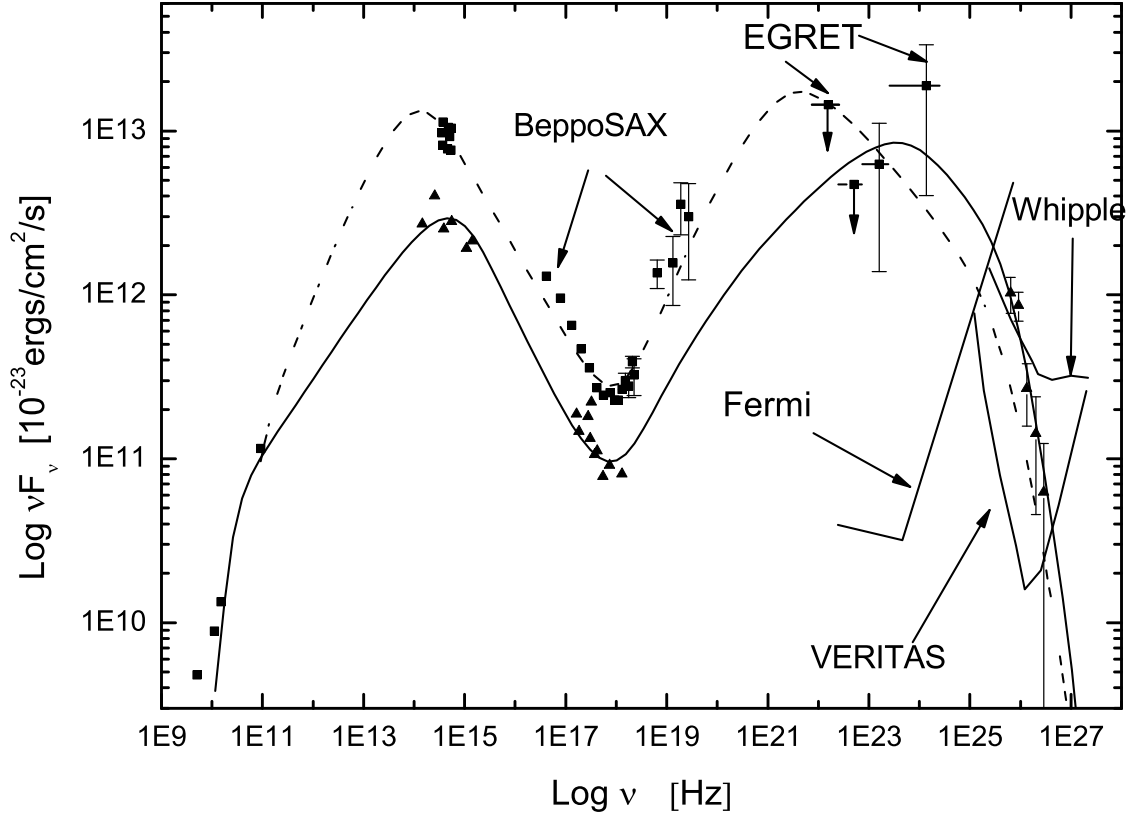}
\caption{Broadband SEDs of the optical/X-ray outburst in 1998
({\em squares}) and of the TeV flare in 2008 ({\em triangles})
with the model fits. The SED data of the optical/X-ray outburst
are taken from Tagliaferri et al. (2000, for the optical and the
X-ray data) and B\"{o}ttcher et al. (2002, for the radio and the
the EGRET data). The observed data of the TeV flare in 2008 are
from Acciari et al. (2008) and the references therein. The model
parameters are $\delta=24.5$, $B=0.01$ G, and $R=6.35\times
10^{16}$ cm for the fit to the SED of the 2008 TeV flare ({\em
solid line}), and  $\delta=19.4$, $B=0.14$ G, and $R=2.1\times
10^{16}$ cm for the fit to the SED of the 1998 optical/X-ray
outburst ({\em  dashed line}). The sensitivity curves for Whipple,
VERITAS, and Fermi are also shown as marked in the figure.}
   \label{Fig:plot1}
   \end{center}
\end{figure}

\section{Conclusion and Discussion}
\label{sect:discussion and conclusions} The broadband SEDs
observed quasi-simultaneously during the TeV flare in 2008 and
during the optical/X-ray outburst phase in 1998 are fitted with
the single-zone synchrotron+SSC model. In the model, the IC
radiation of the external photon fields (EC) is not taken into
account since the CMB/BLR/torus/accrection disk photon fields are
weak comparing with the synchrotron radiation field for a BL Lac
object. The results show that the SEDs can be described by the
synchrotron + SSC leptonic jet model.

The satisfactory fitting requires a large beaming factor, i.e.,
$\delta\sim 25$ and $\delta\sim 20$ for the TeV flare and the
optical/X-ray outburst, respectively, suggesting that both the
optical/X-ray outburst and the TeV flare are from a relativistic
jet. The strength of the magnetic field for the TeV flare is $\sim
14$ times smaller than that of the X-ray/optical outburst, likely
indicating that the emission region is more distant from the core
in 2008 than in 1998, since the magnetic field \emph{B} decreases
along the jet. Electrons can be reaccelerated after leaving the
inner jet, hence the peak frequencies of synchrotron radiation in
the kiloparsec-scale jets could be more than 100 times larger than
those in the inner jets (e.g., Bai \& Lee 2003). The results favor
this idea.

Almost the entire flare in 2008, which lasts four nights, was
recorded by VERITAS. The simultaneous observations in the X-ray
band did not detect any outburst/flare events. An X-ray flare
following the TeV flare after two weeks was observed, with a peak
flux 4 times higher than that observed during the TeV flare. This
fact shows that the TeV flare and the X-ray flare may be not
simultaneous. As shown in Figure 1, the IC component peaks around
$10^{22}$ Hz for the SED during an optical/X-ray outburst, which
is 2 order of magnitudes lower than that for the SED during the
TeV flare. The GeV observation thus is critical to characterize
the SEDs of both the optical/X-ray outburst and the TeV flare. The
fitting results suggest that the optical/X-ray outburst may
accompany a strong MeV/GeV flare. The predicted flux at GeV is
much above the sensitivity of \emph{Fermi}/LAT, and W com is a
selected target for \emph{Fermi}/LAT. The model could be verified
with the \emph{Fermi}/LAT observation.

In the TeV band, the predicted TeV flux by the model for the
optical/X-ray outburst in 1998 is lower than the sensitivity of
Whipple/IACT. This may be the reason why no detection was obtained
with Whipple/IACT in 1998. Although the predicted TeV flux is
marginally over the sensitivity of VERITAS, it is much lower than
the observed TeV flux in 2008. .

\begin{acknowledgements} I appreciate the referee for his valuable suggestions. I thank Jinming Bai,
Liang Chen and Hongtao Liu for their helpful discussions. This
work is supported by the National Natural Science Foundation of
China (Grants 10573030 and 10533050).
\end{acknowledgements}

\label{lastpage}

\end{document}